\documentclass[pre,aps,floats,twocolumn]{revtex4}

\usepackage{graphicx}
\usepackage{dcolumn}
\usepackage{bm}
\usepackage{amsmath}

\begin{document}

\title{Anomalous Heat Conduction in Three-Dimensional Nonlinear Lattices}
\author{Hayato Shiba}
\email{hshiba@scphys.kyoto-u.ac.jp}
\affiliation{Department of Physics, Kyoto University, Kyoto 606-8502, Japan}
\author{Nobuyasu Ito}
\email{ito@ap.t.u-tokyo.ac.jp}
\affiliation{Department of Applied Physics, University of Tokyo, Tokyo 113-8656, Japan}
\date{\today}

\begin{abstract}
Heat conduction in three-dimenisional nonlinear lattice models is studied
using nonequilibrium molecular dynamics simulations.
We employ the Fermi-Pasta-Ulam-$\beta$ model,
in which nonlinearity exists in the interaction of the biquadratic form.
It is confirmed that the thermal conductivity, the ratio of the energy flux to the temperature
gradient, diverges with increasing system size up to $128\times 128\times 256$ lattice sites.
This size corresponds to nanoscopic to mesoscopic scales of approximately 100nm.
From these results, we conjecture that the energy transport in insulators with
perfect crystalline order exhibits anomalous behavior.
The effects of the lattice structure, random impurities, and the natural length in
interactions are also examined.
We find that fcc lattices display stronger divergence than
simple cubic lattices.
When impurity sites of infinitely large mass, which are thus fixed,
are randomly distributed, such divergence vanishes. 
\end{abstract}

\maketitle
\section{Introduction}
It is a long-standing problem to reproduce
irreversible heat conduction phenomena, 
which are described by the Fourier law, $J=-\kappa\nabla T$,
where $\kappa$ denotes the heat conductivity, 
on the basis of time-reversible microscopic dynamics. 
Recent studies have succeeded in reproducing the
Fourier law\cite{Shimada,Murakami,Ogushi,Kaburaki}, 
and other linear nonequilibrium transport phenomena\cite{Ishiwata,Yuge}. 
However, there remains an unsolved problem.
The studies cited above mainly used particle systems,
such as hard spheres and Lennard-Jones particles, for example, 
but historically, nonlinear lattice systems have also been 
studied\cite{LepriR, FPUbeta, FPUalpha, Hatano, FPUother, AokiKusnezov}. 
Shimada {\it et al.} showed that the Fourier law is reproduced in a
three-dimensional (3D) polymer-like lattice with an Fermi-Pasta-Ulam-$\beta$
(FPU-$\beta$) type interaction\cite{Shimada},  
but Shiba {\it et al.} observed the divergence of thermal conductivity for 3 and 
4D FPU-$\beta$ lattices \cite{Shiba}. 
The purpose of this article is to study the energy transport behavior 
in nonlinear lattice systems. 

Let us start with a brief review of the study of heat conduction in nonlinear lattices. 
It is well-known that no temperature gradients are formed if we use
integrable systems, such as harmonic chains\cite{Lebowitz}. 
To realize linear temperature gradients, we need features that
give rise to thermalization processes.
One possibility for generating such a feature is represented by
nonintegrable nonlinear interactions\cite{Peierls}. 
Nonintegrability results in macroscopically irreversible 
processes, even in conserved systems, even though the basic equation is reversible.

In recent years, heat conduction in 1D nonintegrable
chains has been widely investigated\cite{LepriR}. 
In these studies, it was found that the heat conductivity of 
nonlinear lattices depends on the system size as $\kappa\sim N^\alpha$, 
where $N$ denotes the length of the system, 
when the total momentum of the system is conserved. 
Although the value of the exponent $\alpha$ seems to depend on the model, 
it is approximately 0.4 for several models, for example,  
FPU-$\beta$ lattices\cite{FPUbeta}, 
FPU-$\alpha$ lattices\cite{FPUalpha},
diatomic Toda lattices\cite{Hatano},
and 1D binary hard-sphere gases\cite{hsg, CasatiProsen}
exhibit similar power-law behavior. 
This anomalous behavior originates in the power-law decay 
of the energy-flux autocorrelation function, $C(t) = \langle \bm{J}(t)\cdot \bm{J}(0)\rangle$, 
which is the integrand of the Green-Kubo formula\cite{GK, GK2, GK3}, 
\begin{equation}
\kappa = \frac{1}{k_BT^2V}\int dt\ C(t), \label{eq:GK}
\end{equation}
where $\bm{J}(t)$ is the total heat flux and $V$ is the volume of the system. 
In 1D nonlinear lattices, $V=N$ is the chain length. 
The autocorrelation functions of systems with $\alpha\sim 0.4$ decay
approximately as $C(t)\sim t^{-\beta}$ with $\beta\sim 0.6$. 
Phenomenological explanations for this decay exponent have been
proposed using mode-coupling theory\cite{LepriM},
the renormalization group analysis of 1D fluctuating hydrodynamics\cite{Narayan, MaiNarayan}, 
and kinetic theory\cite{kinetic}.

The same types of divergence of $\kappa$ have been
observed in 1D and 2D fluid systems.
For simple classical fluids, it has been well established that the autocorrelation 
function exhibits power-law decay called ``long-time tail''\cite{longtail, Resibois, Ernst, Ernst2, Ernst3, Ernst4}, 
which takes the form $C(t)\sim t^{-d/2}$ for $d\ge 2$. 
This was discovered by Alder and Wainwright\cite{Alder, Alder2} using computer simulations, 
and the study of long-time tail is still ongoing\cite{longtail2, Nishino}. 
In 3D systems, the conductivity converges in the limit 
$N\rightarrow\infty$, as demonstrated by the Green-Kubo formula, eq. (\ref{eq:GK}). 
In 2D systems, it diverges as $\ln N$. 
Such dimensionality dependence has been confirmed by 
computer simulations with hard-particle systems\cite{Murakami, Shimada}, 
Lennard-Jones fluids\cite{Ogushi}, and other systems.
The system dimensionality plays a key role in the behavior 
of transport coefficents in fluid systems.

How does the conductivity of nonlinear lattice systems depend on their dimensionality? 
One may naively expect a dependence that is similar to that in fluids, assuming that a phonon gas 
can be described by hydrodynamic equations of motion. 
But this appears to be a too naive assumption.
Recently, it was shown that the conductivity of a 3D extension of FPU-$\beta$ lattices  
displays divergence and that this divergence is consistent with the power-law decay of the autocorrelation function\cite{Shiba}. 

Although the derivation of the long-time tail from the linearized hydrodynamic equations using the
mode-coupling hypothesis\cite{Ernst, Ernst2, Ernst3, Ernst4,Narayan,MaiNarayan} implies its validity for mesoscopic to macroscopic scales, but not for microscopic scales, particle systems also possess long-time tails on microscopic scales. 
In simple particle systems, distances on the order of ten times the mean-free path are sufficient 
to observe the characteristic size dependence of the conductivity\cite{Shimada,Murakami,Ogushi}. 
In nonlinear lattices\cite{Shiba}, however, it has been found that anomalous divergence 
continues up to systems of at least $64\times 64\times 128$ lattice sites.
This size corresponds to mesoscopic systems.
This means that the energy transport in nanoscopic to mesoscopic crystals is
more complicated than that in fluid systems. 
From these findings, it is clear that further study of this problem 
is interesting both theoretically and with regard to nano technology.
Such anomalous features should contribute to explanation of the anomalous features in 
heat transport observed in experiments and numerical simulations of 
nano scale systems such as carbon nanotubes\cite{Fujii,Shioya,Maruyama}.

Our model and simulation method are presented in the next section. 
The robust divergence of the thermal conductivity in 3D
nonlinear lattices is demonstrated in the following two sections: 
In \S 3, a simple cubic lattice is treated, and in \S 4, 
an fcc lattice is treated. 
Such divergence is also shown to exist in a system with a nonlinear interaction
possessing a natural length scale in \S 5.
In \S 6, the effect of the randomness of the mass is studied.
In \S 7, the results for a 2D nonlinear lattice are given. 
The last section contains a summary and conclusion. 

\section{Model and Simulation Method}

The first model we study is a simple 3D extension of FPU-$\beta$ lattices\cite{Shiba}. 
The model Hamiltonian is 
\begin{equation}
\mathcal{H} = \sum_{i=1}^N \frac{\bm{p}_i^2}{2m} + \sum_{\langle i,j\rangle} \left[ \frac{k}{2}|\bm{r}_i -\bm{r}_j|^2 +\frac{g}{4}|\bm{r}_i -\bm{r}_j|^4\right], \label{eq:3dhamil}
\end{equation}
where the 3D vectors $\bm{p}_i$ and $\bm{r}_i$ denote
the momentum and displacement of a particle at lattice point $i$, respectively, and 
the mass $m$ is taken to be unity. 
The summation over $\langle i,j\rangle$ is a sum over nearest neighbors.
$k$ and $g$ are parameters indicating the strength of interactions between these 
nearest-neighbor particles, and they are fixed as $k=1.0$ and $g=0.1$ in this paper.

We point out here that the spatial degrees of freedom of these dynamical models consist of 
the displacement vector $\bm{r}_i = \bm{q}_i -\bm{q}_i^0$, where $\bm{q}_i$ represents 
the real spatial coordinates of the particles, and $\bm{q}_i^0$ represents the equilibrium positions
of the particles, which are at simple cubic lattice points. In other words, 
this model system possesses a crystal structure. 
In this model the longitudinal and transverse modes are treated identically,
and thus have the same dispersion relation in the harmonic limit.

The system size is denoted by $N_x\times N_y\times N_z$. 
We express the system size  in terms of the number of particles $(N_x, N_y, N_z)$, 
not the lengths $(L_x, L_y, L_z)$. Thus, the system size is 
dimensionless $N_c=L_c/a\ (c=x,y,z)$. Here, $a$ is the lattice spacing 
constant, which does not appear in the Hamiltonian given by eq. (\ref{eq:3dhamil}).
The only characteristic length scale that appears in eq. (\ref{eq:3dhamil}) is 
$\sqrt{k/g}$, which is a typical length of phonon-phonon interactions. 

We use periodic boundary conditions in the $x$- and $y$- directions.
The particles on both ends in the $z$-direction are attached to rigid walls, 
which are separated by one lattice space from the walls.
These particles interact with the wall through a potential that is 
identical to that for particle-particle interactions, 
and their local temperature is controlled by the Nos\'e-Hoover method\cite{Nose}.
Thus, the equations of motion are modified for these particles as
\begin{equation}
\dot{\bm{r}}_i = \frac{\bm{p}_i}{m},\quad \dot{\bm{p}_i} =-\frac{\partial\mathcal{H}}{\partial\bm{r}_i} -\zeta_i \bm{p}_i.
\end{equation}
Here, $\zeta_i$ denotes the Nos\'e-Hoover thermostat variables, which obey
\begin{equation}
\dot{\zeta_i} = \frac{1}{Q}\left( \frac{\bm{p}_i^2}{3mk_BT_i}-1\right),
\end{equation}
where $T_i$ denotes the temperatures of the heat baths, which are set to $T_{\rm L}$ on the left
and $T_{\rm R}$ on the right. (Here, $i_z$ increases to the right.)
The difference between $T_{\rm L}$ and $T_{\rm R}$ drives the energy transport in the system. 
The quantity $Q$ is the relaxation time of the heat baths, which is set to unity. 

For particles in the bulk, the equations of motion are
\begin{equation}
\dot{\bm{r}}_i =\frac{\bm{p}_i}{m},\quad \dot{\bm{p}_i}= -\frac{\partial\mathcal{H}}{\partial\bm{r}_i}.
\end{equation}
Particle dynamics simulations were used to study this system. 
The initial displacements were zero for all particles ({\it i.e.}, $\bm{r}_i=\bm{0}$) 
and the initial momenta $\bm{p}_i$ were randomly assigned for each particle. 
Starting from this initial state, initial relaxation steps were discarded. 
After the system reached a steady state, the temperature distribution, 
energy flux, and thermal conductivity were computed. 

\paragraph{Temperature}
Using the Virial theorem, we define the local temperature $T(i)=T(i_x, i_y, i_z)$ 
of a particle on lattice point $i$ as the long-time average of the kinetic energy.
(Here, $i_x$, $i_y$. and $i_z$ are the labels of the lattice sites, 
with $1\le i_x\le N_x,\quad 1\le i_y\le N_y,$ and $1\le i_z\le N_z$.)
\begin{equation}
\frac{3}{2}k_BT(i) = \overline{\frac{\bm{p}_i^2}{2m}}.
\end{equation}
To obtain better accuracy, we averaged the temperature over the 
$N_x\times N_y$ particles in the same cross section, and thus define $T(i_z)$ as
\begin{equation}
\frac{3}{2}k_BT(i_z) = \frac{1}{N_xN_y} \sum_{i_x,i_y} T(i_x, i_y, i_z).
\end{equation}

\paragraph{Heat flux}
In this paper, we refer to energy flux as the heat flux, 
because the system itself is a thermodynamic system that relaxes 
to equilibrium due to its nonintegrability. 
Because the heat flux is conserved in the bulk region, there are two ways to measure it.
One way is to sum up the local contributions to the energy flow due to the interactions. 
The microscopic energy transfer from site $i$ to site $k$ is given by
\begin{equation}
j_{i\rightarrow k} =\frac{1}{2}\left( \frac{\bm{p}_i}{m} +\frac{\bm{p}_k}{m}\right)\cdot \frac{\partial V_{ik}}{\partial\bm{r}_i},
\end{equation}
where $V_{ik}$ is the interaction potential between sites $i$ and $k$. 
By performing this summation, we could obtain the total heat flow density
as $\langle j_z\rangle = N^{-1}\sum_{\langle j,k\rangle} j_{i\rightarrow k}$. 
However, in this study, we used an alternative method;
we calculated the work done by the heat baths as 
\begin{equation}
\langle j_z\rangle = \frac{1}{N_xN_y}\left|\sum_{i\in\textrm{baths}} \zeta_i\bm{p}_i\cdot \frac{\bm{p}_i}{m} \right|.
\end{equation}
Here, the summation is over all the particles to which the left (or right)
heat baths are attached. We have confirmed that these two methods 
yield the same value of the heat flow to at least two significant figures.

\paragraph{Thermal conductivity}
In the simulations carried out for this study,
the temperature distribution was found to be linear in the system 
and no temperature gap was observed. 
Therefore, the thermal conductivity $\kappa (N_z)$ can be estimated by 
\begin{equation}
\kappa (N_z) = \frac{\langle j_z\rangle N_z}{T_L-T_R}. \label{eq:conductivity}
\end{equation}

For the numerical integration of the equations of motion, 
we used a St\"ormer-Verlet difference scheme, given by
\begin{eqnarray*}
p_i^{n+1/2} &=& p_i^n -\frac{\Delta t}{2}\left[ \frac{\partial}{\partial q_i}V(\{ q^n\} ) +\zeta_i^n p_i^{n+1/2} \right], \\
q_i^{n+1} &=& q_i^n + \Delta t\ \frac{p_i^{n+1/2}}{m}, \\
\zeta_i^{n+1} &=& \frac{\Delta t}{Q} \left[ \sum_i \frac{(p_i^{n+1/2})^2}{3mk_BT} -1 \right] \\
\mbox{and} & & \\
p_i^{n+1} &=& p_i^{n+1/2} -\frac{\Delta t}{2}\left[ \frac{\partial}{\partial q_i}V(\{ q^{n+1} \}) +\zeta_i^{n+1} p_i^{n+1/2} \right].
\end{eqnarray*}

Here, $n$ represents the number of time steps. 
This scheme exhibits second-order convergence with $\Delta t$. 
For the bulk particles, without temperature control, 
this scheme reduces to the leap-frog integrator, which is symplectic. 
Therefore, the integration of the bulk area is numerically stable 
for all $\Delta t$ satisfying $\Delta t\le 0.02$. 
However, this scheme is not symplectic for the thermostatted particles. 
The time step $\Delta t$ is taken to be small enough
for the integration of these particles connected to the heat bath to be stable.
Throughout this paper, $\Delta t$ is set to 0.02.

\section{Simple Cubic Lattice}

The results for the simple cubic FPU-$\beta$ lattice are given 
in this section up to $N_z=256$.
(A previous study gave results up to $128$\cite{Shiba}.)
For each data point, the results are averaged over 
simulations starting from five independent initial conditions.
 
\begin{figure}
\centerline{
\resizebox{0.48\textwidth}{!}{\includegraphics{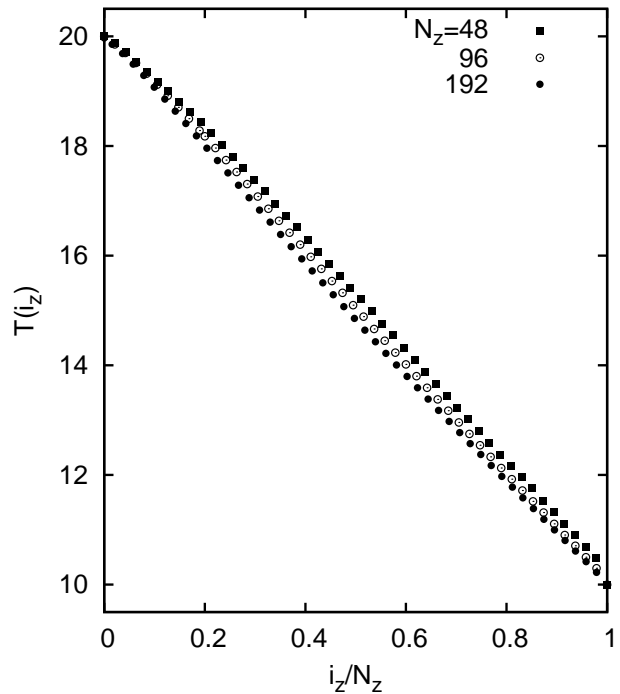}}}
\caption{Temperature profile for the FPU-$\beta$ model with a simple cubic lattice.
The sequences represent the results for lattices with $N_z=$48, 96, and 192, from top to bottom. 
The horizontal axis represents the position along the $z$-direction, 
rescaled by the system size $N_z$. The local temperature is averaged 
over a cross-sectional cut in the $xy$-plane. 
The $3\sigma$ width is smaller than the symbols.}
\label{fig:Temp1}
\end{figure}

\begin{figure*}
 \centerline{\resizebox{0.5\linewidth}{!}{\includegraphics{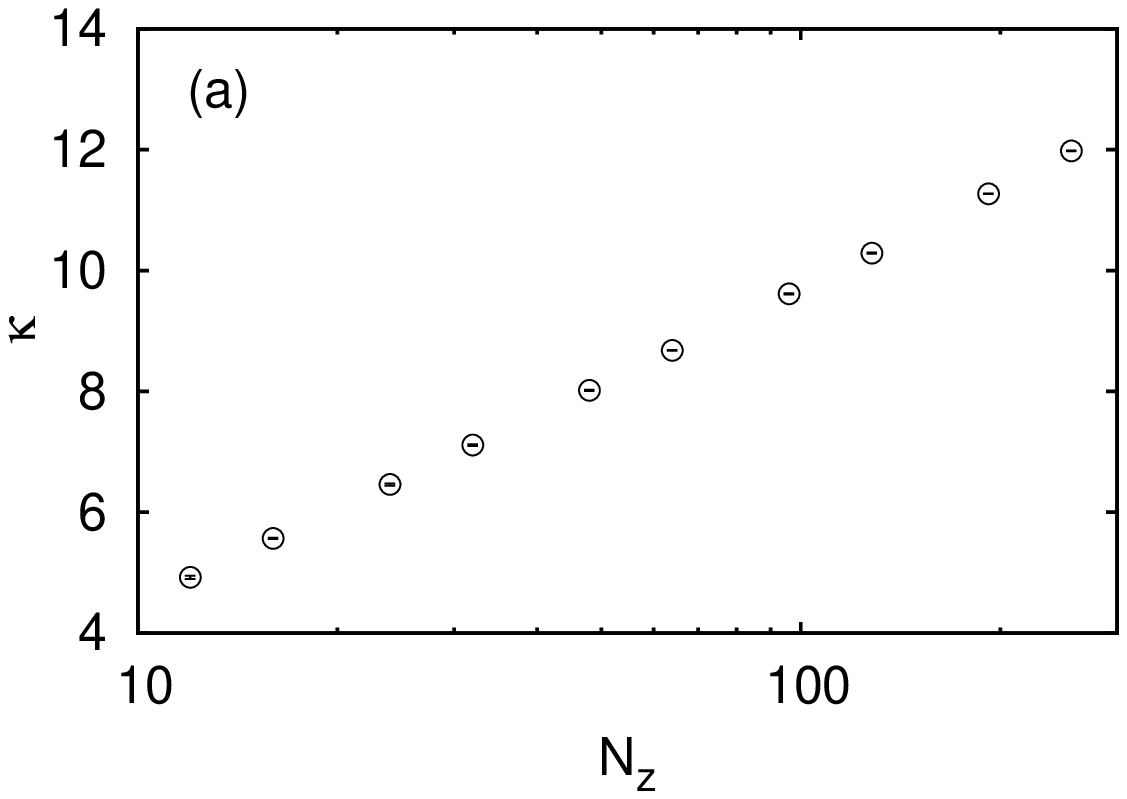}} \quad \resizebox{0.5\linewidth}{!}{\includegraphics{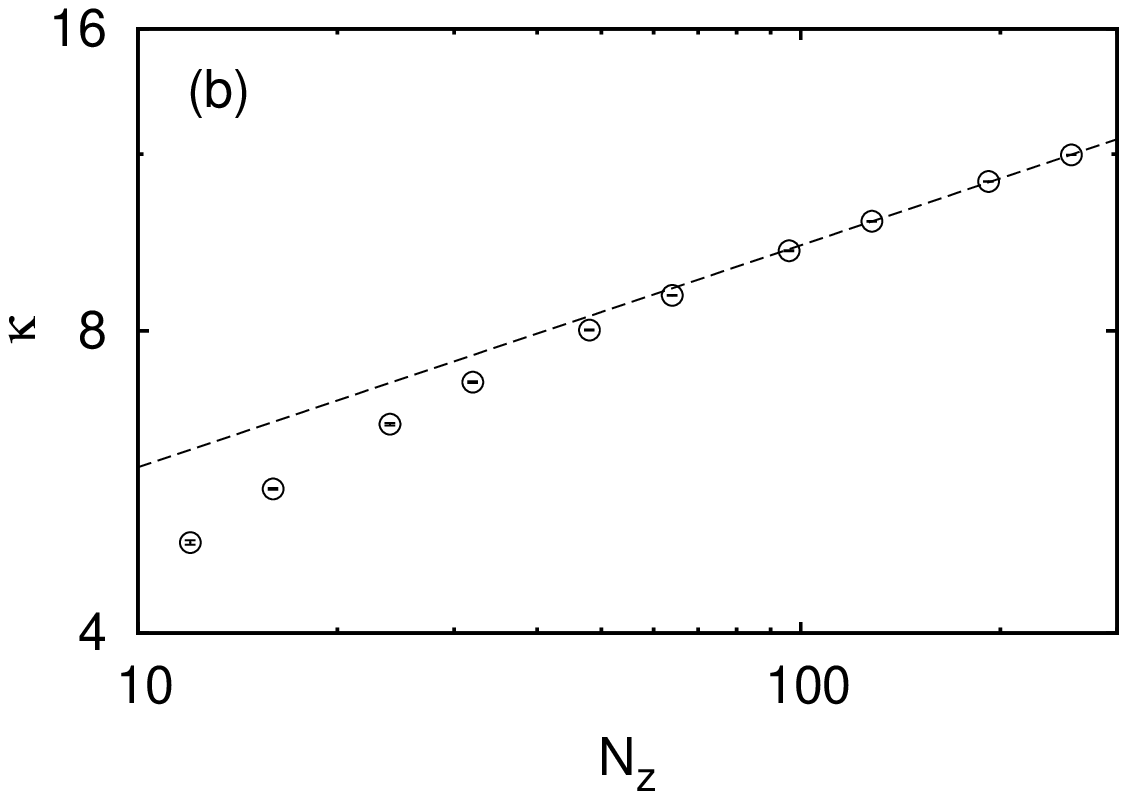}}}
\caption{\label{fig:FPU3D} (a) System size dependence of the thermal conductivity 
for the simple cubic FPU-$\beta$ model plotted on a semi-log scale. 
(b) The same data plotted on a log-log scale. A power-law fitting was carried out
for the data with $N_z\ge 96$. The result is $\kappa\sim N_z^{0.221(4)}$,
which is plotted by on a broken line. 
}
\end{figure*}
First, we present the results for the temperatures $(T_L, T_R) = (20.0, 10.0)$. 
For this case, the spatial temperature profile is plotted in Fig. \ref{fig:Temp1} 
for systems with $N_z=48$, $96$, and $192$ 
with the positions of the particles rescaled by the system size $N_z$.  
No boundary gap is observed in this temperature profile, 
and the results $T(i_z/N_z)$ for all values of $N_z$ nearly coincide.
We thus conclude that we can estimate the thermal conductivity 
using eq. (\ref{eq:conductivity}).  The system size dependence 
of the estimated values of $\kappa (N_z)$ are plotted in Fig. \ref{fig:FPU3D} 
on semi-log and log-log scales, up to the system size $128\times 128\times 256$. 
These results for $\kappa (N_z)$ suggest logarithmic divergence,
although power-law divergence is not excluded. 
When a power law  
is fitted to the data for $96\le N_z\le  256$, we obtain $\kappa\sim N_z^{0.221(4)}$.
Such a divergence was previously observed up to $N_z=128$\cite{Shiba}, 
and it is now confirmed up to $N_z=256$. 
Note that the results show no sign of convergence.

\begin{figure}[b]
\centerline{
\resizebox{0.5\textwidth}{!}{\includegraphics{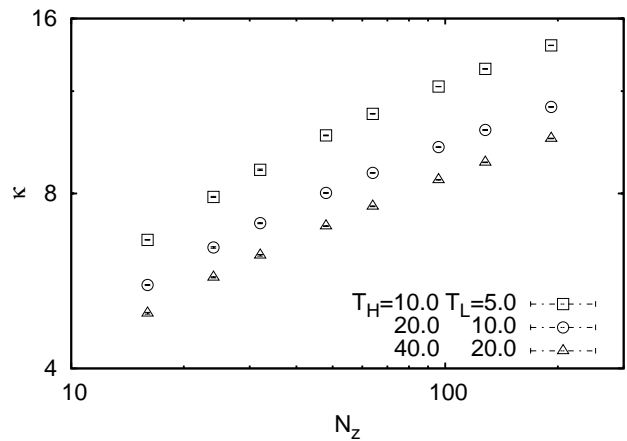}}}
\caption{System size dependence of the thermal conductivity for the temperatures $(T_L, T_R)=(10.0, 5.0), (20.0, 10.0),$ and $(40.0, 20.0)$. The data for $(T_L, T_R)=(20.0, 10.0)$ are the same as those in Fig. \ref{fig:FPU3D}}.  
\label{fig:Tdiff2}
\end{figure}

The behavior of $\kappa (N_z)$ when the heat bath 
temperatures are changed is shown in Fig. \ref{fig:Tdiff2}. 
For the same $N_z$, thermal conductivity becomes lower for higher $(T_L,T_R)$.
This is expected because thermalization is enhanced when the temperature
is high because of the stronger nonlinear interactions. 
However, we observe that the divergence of $\kappa (N_z)$ with increasing
system size is observed even at higher temperatures such as
$(T_L, T_R) = (40.0, 20.0)$. 

After beginning the simulation with the initial conditions
described in \S 2, we waited for the time period $t_w \sim 10^5$. 
By this time, the system had relaxed to a nonequilibrium steady state, 
where the amount of energy flow per time was stationary.
For the $128\times 128\times 256$ system, 
which is the largest system considered in the present study,
and includes about $4\times 10^6$ particles, the total number of simulation steps 
multiplied by the number of particles is of the order of $10^{13}$. 
For the $128\times 128\times 256$ system, 
approximately 20h of CPU time was required
to numerically calculate one sample,
using an SX8 cluster with a peak performance of 128 GFlops. 

\section{FCC Lattice}
\begin{figure*}
\centerline{
  \resizebox{0.5\linewidth}{!}{\includegraphics{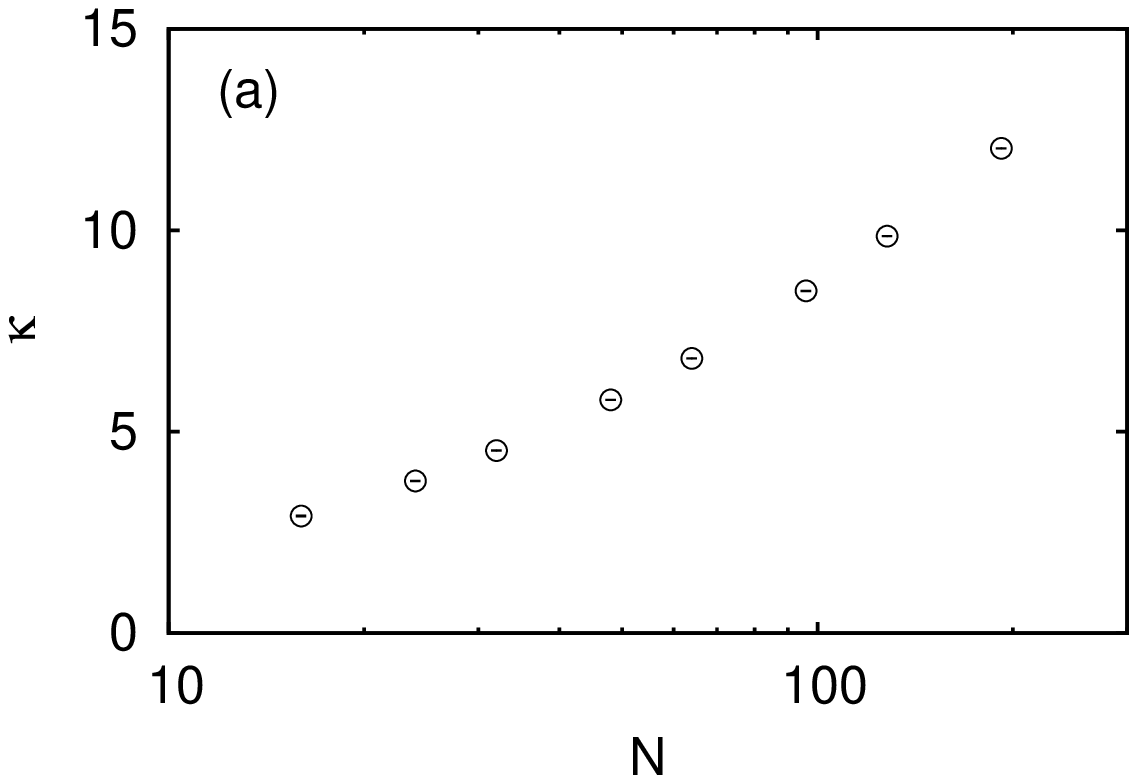}} \quad \resizebox{0.5\linewidth}{!}{\includegraphics{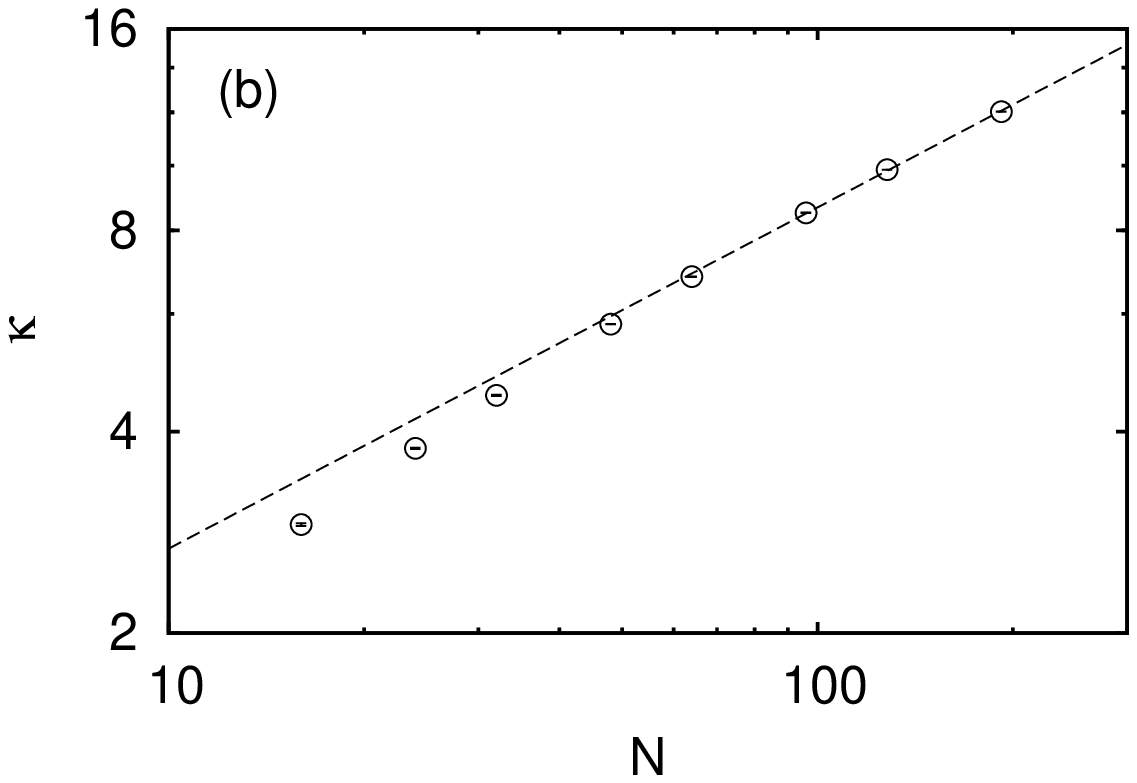}}}
\caption{Size dependence of thermal conductivity of an fcc FPU-$\beta$ nonlinear lattice 
on (a) semi-log and (b) log-log scales. The fit of the result for $N\ge 64$ is plotted
with the dotted line, which represents $\kappa \sim N^{0.51(1)}$.
} 
\label{fig:FCC2}
\end{figure*}
We also studied the thermal conductivity of an fcc system
in order to determine its dependence on the lattice structure.
The Hamiltonian of this model is also that given in eq. (\ref{eq:3dhamil}), 
but in this case, each particle interacts with its twelve nearest neighbors. 
Lattices of sizes $N\times N\times (N+1)$ are used, and each lattice point is 
again denoted by $(i_x, i_y, i_z)$. 
In these simulations, particles only occupy sites where $(i_x+i_y+i_z)$ is even. 
Therefore, the total number of particles is $N^2(N+1)/2$. 
Nos\'e-Hoover heat baths were attached to the 1st and $(N+1)$th layers in the $z$-direction. 
In these layers, the particles are connected to the walls with the same interactions as those
in the system discussed above (fixed boundary condition). 
Periodic boundary conditions were adopted in the $x$- and $y$-directions. 
Similarly to the simple cubic case, It takes a time period of less than $10^5$ for 
the system to reach a nonequilibrium steady state. 

The estimated values of the thermal conductivity are shown in Fig. \ref{fig:FCC2} 
up to the system size of $192\times 192\times 193$. 
The temperatures of the heat baths here are  $(T_L, T_R)=(20.0, 10.0)$. 
In this figure, power-law divergence is clearly observed, and in this case, 
the possibility of logarithmic divergence is excluded.
Fitting in the region $64\le N\le 192$ yields $\kappa (N)\sim N^{0.51(1)}$. 
This value of the exponent is larger than that for the simple cubic lattice 0.221(4). 
This suggests that the power-law exponent of the divergence is not determined 
only by the lattice dimensionality. 
It is interesting that this divergence seems to be stronger than that in 1D models,
for which the exponent was found to be approximately $0.4$\cite{LepriR}.


\section{Natural Length}
Using the Hamiltonian in eq. (\ref{eq:3dhamil}), we cannot realize a system 
in which longitudinal and transverse modes are treated differently. 
However, in real solids, the dispersion relations for these modes generally differ, 
and they can have very different structures. 
To include such effects, we need to consider the interactions among the particles 
more carefully. 
\begin{figure}[b]
\centerline{
\resizebox{0.5\textwidth}{!}{\includegraphics{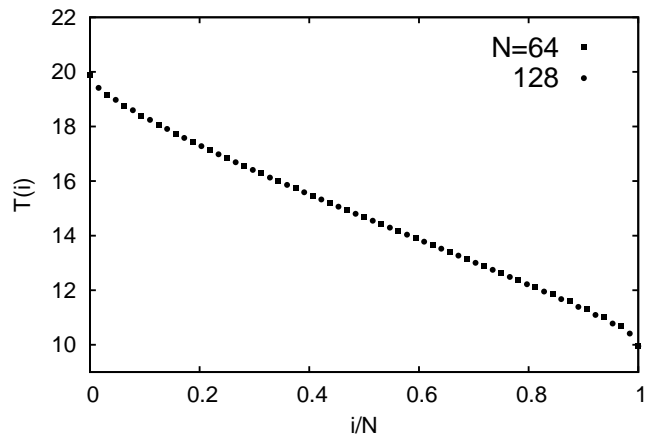}}}
\caption{Temperature profiles of the fcc nonlinear lattice whose Hamilitonian
 is given by eq. (\ref{eq:NL}). Results for the sizes $N=64$ and 128 are displayed.
 Their curves are indistinguishable in this plot.
 }
\label{fig:tempfccn}
\end{figure}

Here, we consider the inclusion of the natural length of the interaction
as a first step, and we consider the modified Hamiltonian
\begin{equation}
\mathcal{H} = \sum_{i=1}^N \frac{\bm{p}_i^2}{2m} +\sum_{\langle i,j\rangle} \left[ \frac{k}{2}(dq_{ij})^2 +\frac{g}{4}(dq_{ij})^4 \right], \label{eq:NL}
\end{equation}
where $dq_{ij} = |\bm{q}_i -\bm{q}_j| -l_0$, and $l_0$ is the natural length of the interaction. 
A system comprising a simple cubic lattice was investigated using this Hamiltonian
by Shimada {\it et al.}, and they observed normal heat conduction in three dimensions\cite{Shimada}. 
However, this model has the flaw that its transverse modes are softened, 
and as a result, the crystal structure was not maintained in their simulations.

\begin{figure*}
\centerline{
  \resizebox{0.5\linewidth}{!}{\includegraphics{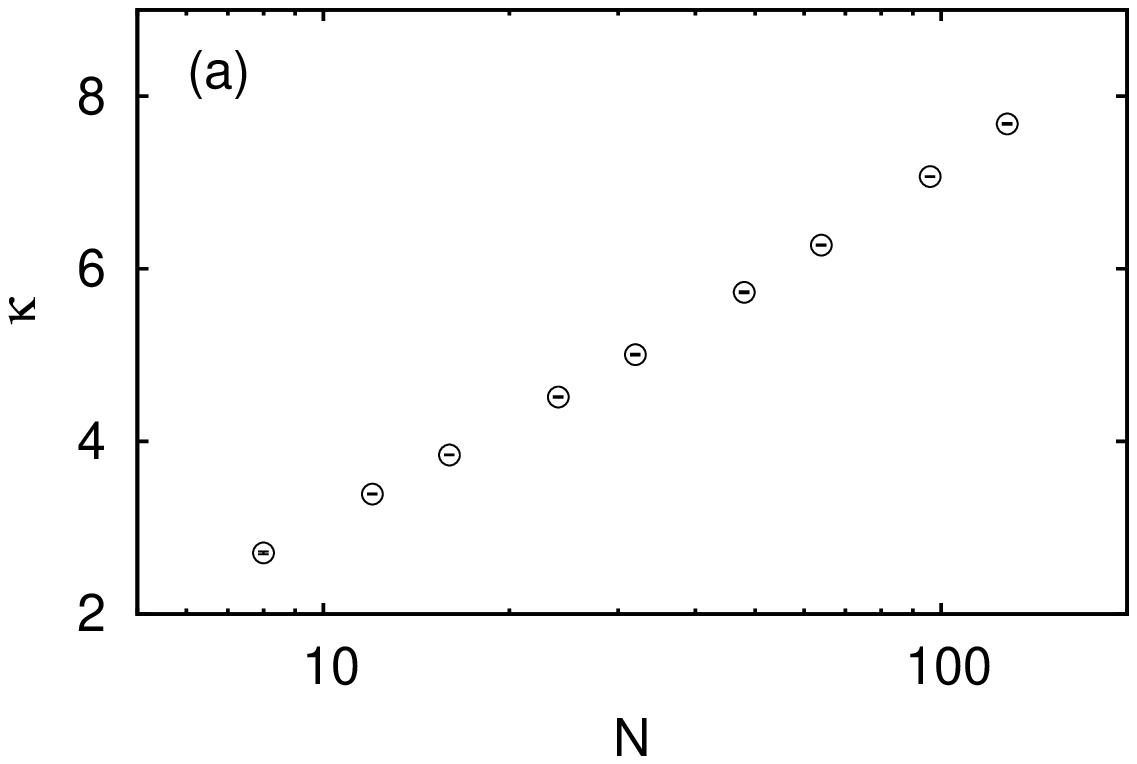}} \quad \resizebox{0.5\linewidth}{!}{\includegraphics{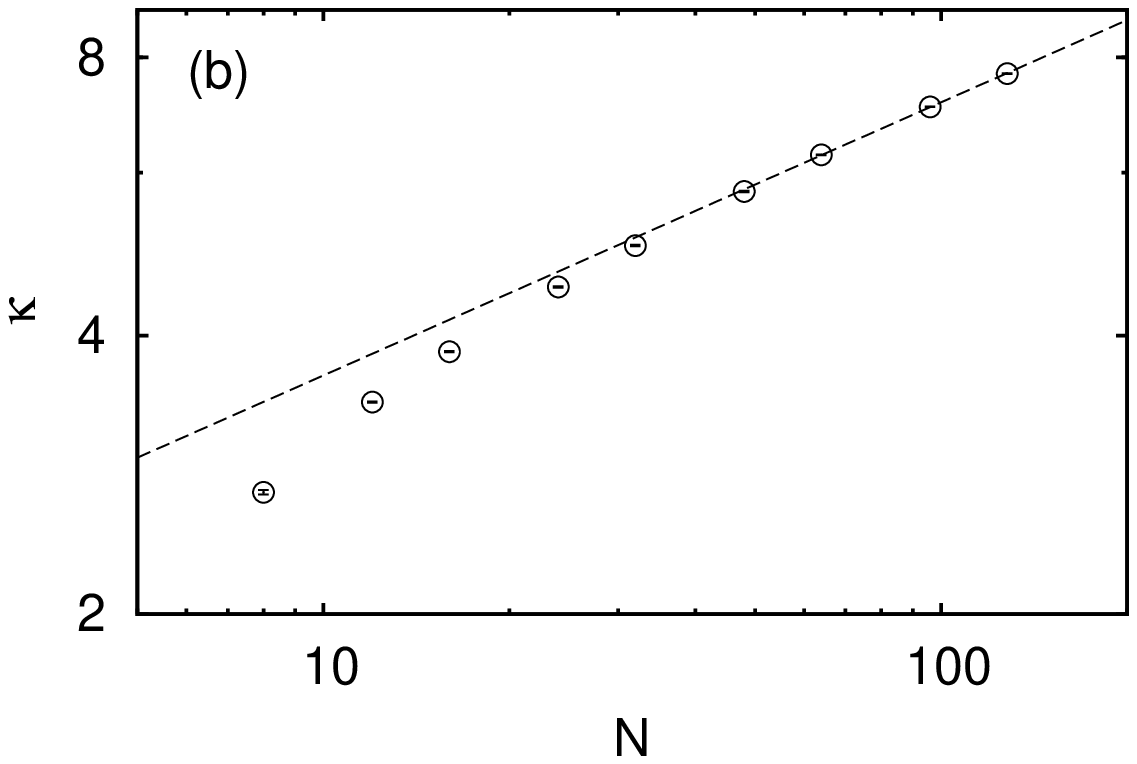}}}
\caption{System size dependence of the thermal conductivity for an fcc lattice whose
Hamiltonian is given by eq. (\ref{eq:NL}). On the left is a plot on a semi-log scale, 
and on the right is a plot of the same data on a log-log scale. 
In the right figure, the dotted line represents the result of a power-law fitting 
for the region $N\ge 48$ giving the result $\kappa\sim N^{0.295(4)}$.
}
\label{fig:fccn}
\end{figure*}
In order to realize a system that maintains a crystal structure in thermal equilibrium 
employing such a class of Hamiltonians, we need to employ a stable structure. 
One possibility is the fcc lattice, which can maintain its structure
with only nearest-neighbor interactions. In this section, we present 
simulation results for systems comprising fcc lattices, 
employing the Hamiltonian in eq. (\ref{eq:NL}).
Here, we set the parameters as
\begin{equation}
k=1.0,\ g=0.1,\ l_0=50.0\times \sqrt{2}.
\end{equation}
The lattice structure here is the same as that in the previous sections. 
Thus, the heat flux flows along one of the face-centered directions, 
and the lattice constant is $\sqrt{2}l_0 = 100.0$. The temperatures of the
heat baths were also set to $(T_L, T_R)=(20.0, 10.0)$ in this case. 

The temperature profiles are plotted in Fig. \ref{fig:tempfccn}. 
We again observe that the temperature gaps near the walls are small and that
the results $T(i_z/N_z)$ are the same for different system sizes $N$.
We can therefore obtain the system size dependence of thermal conductivity
using eq. (\ref{eq:conductivity}). 
As in the previous case, we consider $N\times N\times (N+1)$ lattices with
$N^2 (N+1)/2$ particles. 

The system size dependence of thermal conductivity is plotted in Fig. \ref{fig:fccn}
on both semi-log and log-log scales. It is clearly seen that the divergence is not
logarithmic but of a power-law type. Fitting the data for the region $N\ge 48$, 
we obtain a tentative value for the power-law exponent of $\alpha\sim 0.295(4)$. 

This result suggests that the anomalous behavior we have observed is not 
peculiar to a Hamiltonian of the form of eq. (\ref{eq:3dhamil}), and 
we conjecture that it is a general property of insulating solids. 
So far, all the results we have obtained for nonlinear lattice models
with perfect crystalline order are consistent with anomalous thermal conductivity.
It is a future problem to investigate the extent to which this 
anomalous behavior can be observed.

\section{Disorder}\subsection{Random mass}
The purpose of this paper is to show that the divergence of thermal conductivity 
is a robust property of nonlinear lattices, even in 3D cases. 
It has been a long-standing belief that the following 
play important roles in thermal and transport phenomena:
(i) nonlinearity in the interactions, (ii) the dimension of the system, and (iii) impurities or disorders. 
In the previous sections, we showed that anomalous behavior exists
if we include only (i) and (ii). In this section, we investigate some systems with disorder.

In 1D cases, disordered harmonic chains have been investigated 
for many years\cite{HDO1, HDO2, HDO3, HDO4}. 
For such systems driven by a Langevin thermostat, it has been proved 
that there is a unique steady state with $\kappa (N)\sim N^{1/2}$\cite{HDO2, HDO3, HDO4}. 

Li {\it et al.} investigated heat transport in disordered FPU chains using Nos\'e-Hoover 
thermostats\cite{LiD}. They investigated heat transport at various temperatures and 
found that for sufficiently high temperatures, 
the thermal conductivity diverges as $\kappa (N)\sim N^{0.43}$. 
Although in the case of Nos\'e-Hoover thermostats, at low temperatures
a unique nonequilibrium steady state might not exist,  
when the temperature $T$ is much higher than 0.5,
there should be no such problem.
\begin{figure*}
\centerline{
  \resizebox{0.5\linewidth}{!}{\includegraphics{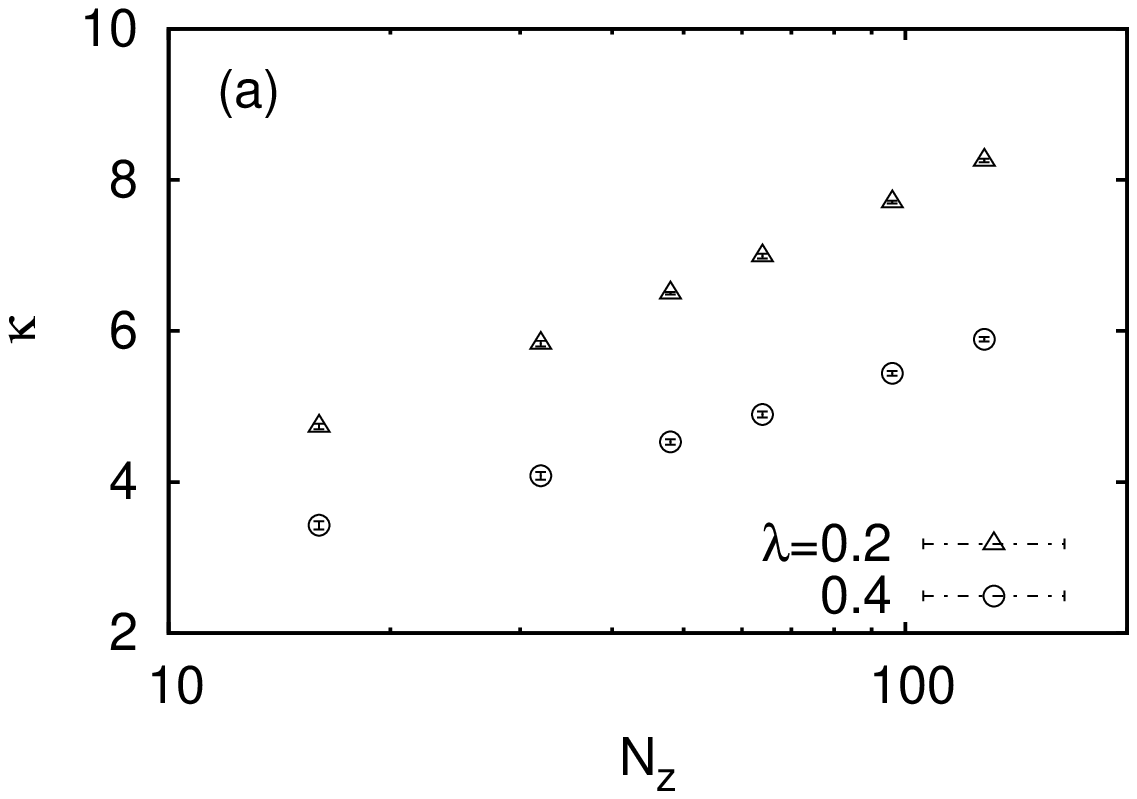}} \quad   \resizebox{0.5\linewidth}{!}{\includegraphics{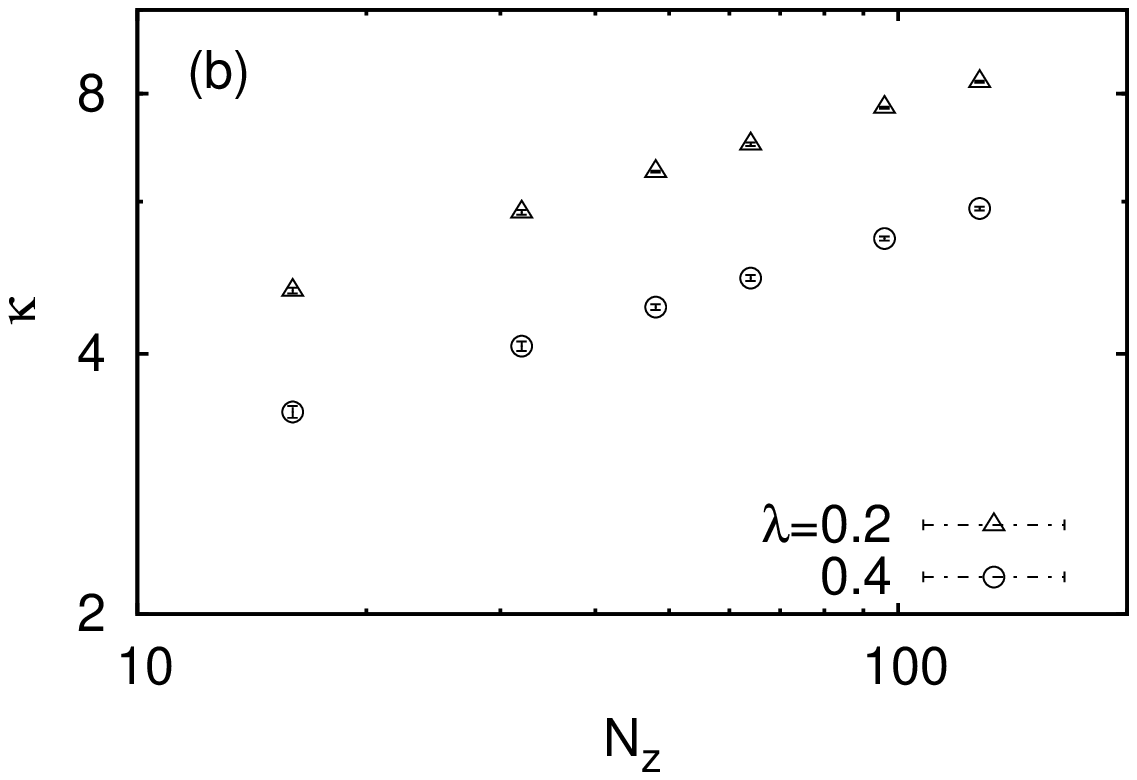}}}
\caption{(a) System size dependence of thermal conductivity of 3D disordered FPU-$\beta$ lattices with the Hamiltonian given by eq. (\ref{eq:massrand}), plotted on semi-log scale. The cases of two strengths of the disorder parameter $\lambda =0.2$ and $0.4$ are investigated. (b) Plot of the same data on log-log scale. \\ \\
\label{fig:massrand}
}
\end{figure*}
To eliminate this possible problem, therefore, we only investigated 
3D systems at high temperatures. 
The model we investigated has the following Hamiltonian:
\begin{equation}
\mathcal{H} = \sum_{i=1}^N \frac{\bm{p}_i^2}{2m_i} +\sum_{\langle i,j\rangle} \left[ \frac{k}{2}|\bm{r}_i -\bm{r}_j|^2 +\frac{g}{4} |\bm{r}_i -\bm{r}_j|^4\right]. \label{eq:massrand}
\end{equation}
The only difference between this Hamiltonian and that appearing in eq. (\ref{eq:3dhamil}) 
is that, in the above form, the masses of the particles $m_i$ vary among lattice sites. 
This allows us to include disorder in the form of random masses.
We chose each $m_i$ as a random number given by
\begin{equation}
m_i = m_0 + \lambda (R_i -0.5),
\end{equation}
where $\lambda$ is a parameter adjusting the amplitude of randomness, and $R_i$ is a random number distributed uniformly on the interval of $[0,1)$. We set the average mass $m_0$ to unity.  

Using the Hamiltonian in eq. (\ref{eq:massrand}), we carried out simulations for a system with a
simple cubic lattice. The setup for the simulation was similar to that for the simulations discussed
in \S 3, with $N_x:N_y:N_z=1:1:2$. The averages of the thermal conductivities obtained in 5
mass configurations were computed for various system sizes and two values of $\lambda$. 
The system size dependence of the thermal conductivity is plotted in Fig. \ref{fig:massrand}. 
It is observed that the thermal conductivity is lower when the effect of the disorder 
is stronger and that, here too, thermal conductivity diverges with increasing system size. 
We conclude that disorder does not destroy the anomalous thermal conductivity. 

\subsection{Randomly fixed sites}
Because we could find no indication of normal heat conduction
in the simulations considered so far, 
we also studied one extreme situation when the disorder effect 
should be very strong, that in which the system possesses impurities of infinite mass. 
Although the form of the Hamiltonian in this case is also given by eq. (\ref{eq:massrand}), 
the mass was chosen as 
\begin{equation}
m_i = 1\ \textrm{or}\ \infty,
\end{equation}
with a certain fraction of the masses set to infinity.
In other words, we fixed a certain fraction of the particles at their
mechanical equilibrium position with $\bm{r}_i=\bm{0}$. 
We selected the fixed particles randomly from all the particles that
are not connected to the heat baths with proportions of 10\% and 20\%. 
These values are both below the critical percentage above which
the paths of the heat flow are completely blocked. 
\begin{figure*}
\begin{center}
	\resizebox{0.48\linewidth}{!}{\includegraphics{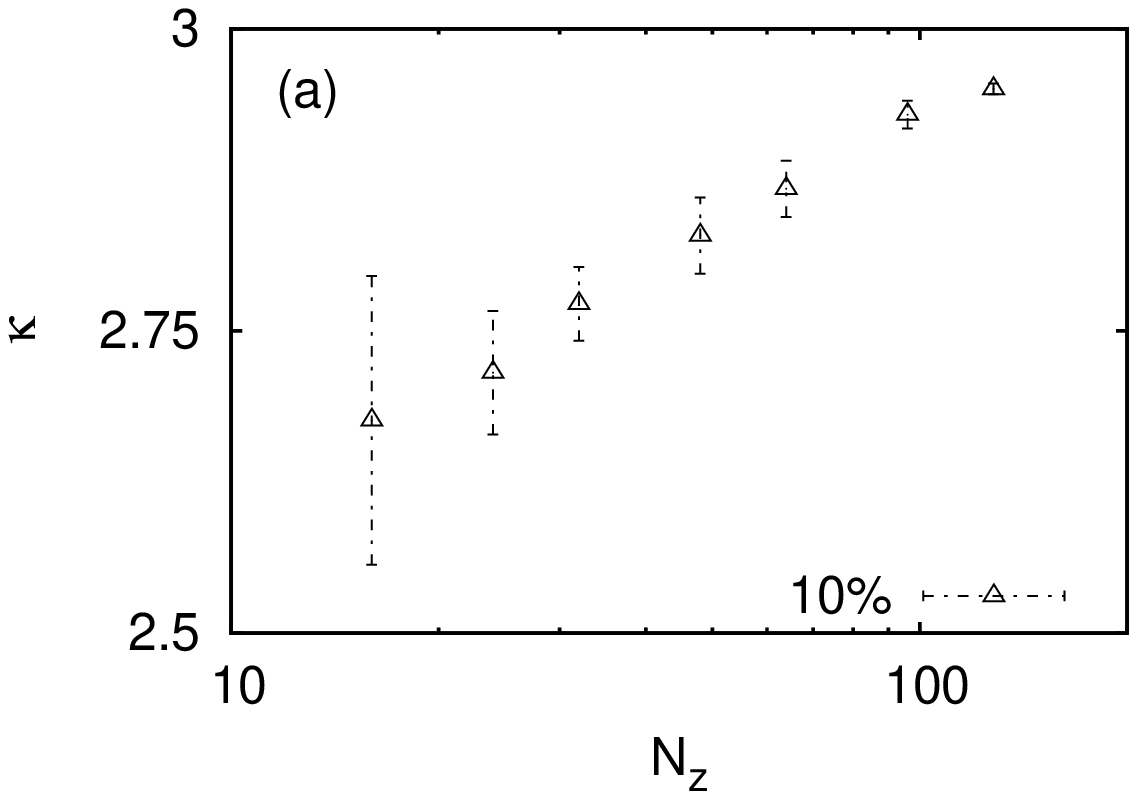}} \
	\resizebox{0.48\linewidth}{!}{\includegraphics{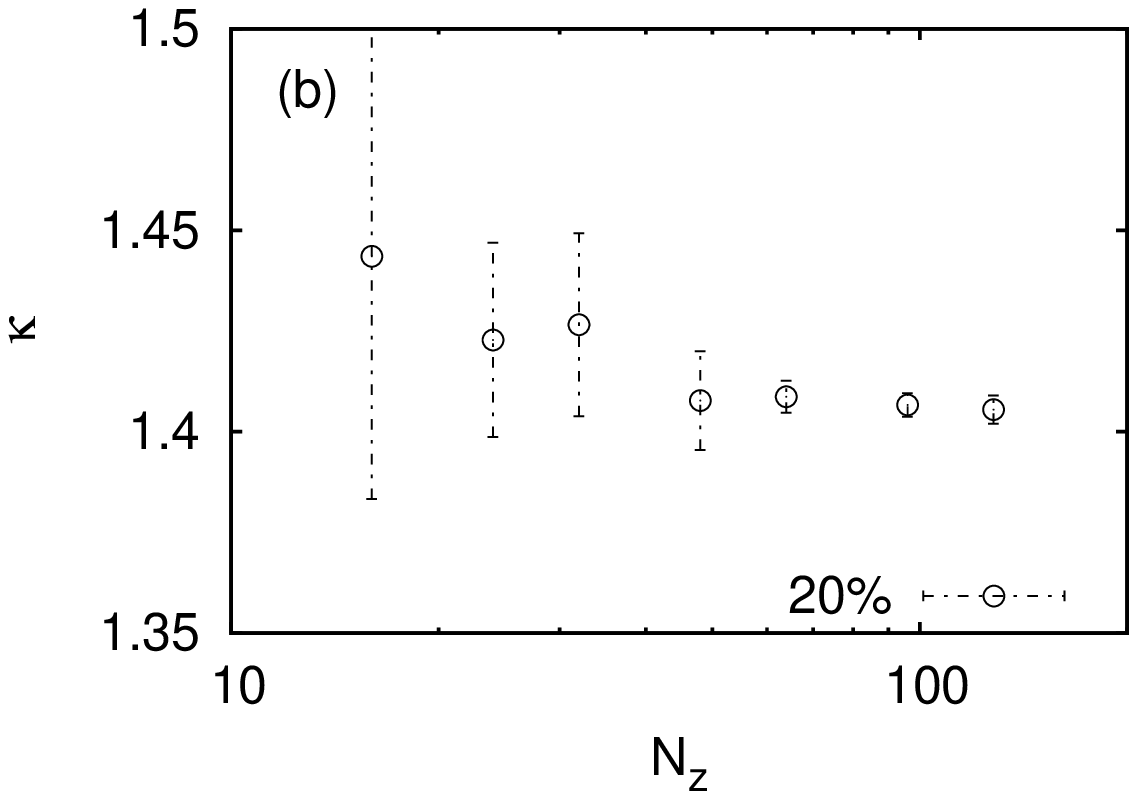}}
\caption{ (a) System size dependence of the thermal conductivity for a model with fixed sites. 
The fraction of fixed sites is 10\%. 
(b) As (a), but with 20\% fixed sites. 
It is seen that the conductivity approaches a constant value with increasing $N_z$. 
} 
\label{fig:fixed}
\end{center}
\end{figure*}

The system size dependence of the thermal conductivity is plotted in Fig. \ref{fig:fixed}. 
It is seen that although the model with $10\%$ fixed sites exhibits some divergence, 
the thermal conductivity of the model with $20\%$ fixed sites clearly converges to 
a finite value as $N_z$ is increased. 
These results suggest that defects in crystalline solids play an important role 
in thermal conductivity behavior. 
Moreover, the conductivity clearly depends on the density of the fixed sites. 

We now remark on past works investigating the 2D harmonic model 
with missing bond defects\cite{mbd} (not missing particles), in which
each particle has a scalar (not vector) dynamical variable, $(p_i, q_i)$. 
This model exhibits convergent thermal conductivity when there are 
sufficiently many missing bond defects, similarly to the system studied here.
More accurate, quantitative, and detailed investigations of the role 
of the various type of defects will be studied in the future. 

\section{Two-Dimensional Lattice}
Here, we comment on some investigations of 2D systems. 
We have already presented evidence that a 3D nonlinear lattice 
system generally exhibits power-law dependence on its system size.
Because it is unlikely that the thermal conductivity of a 2D system 
has weaker divergence than that of a three-dimensional system,
we conjecture that 2D nonlinear lattice systems also 
exhibit power-law divergence of the thermal conductivity.

There have not been many studies on 2D systems.  
Although it is not yet proven, it is widely believed that the thermal conductivity
for such systems diverges logarithmically as the size increases\cite{LepriR, LippiLivi, Yang}.
\begin{figure}
 \resizebox{0.90\linewidth}{!}{\includegraphics{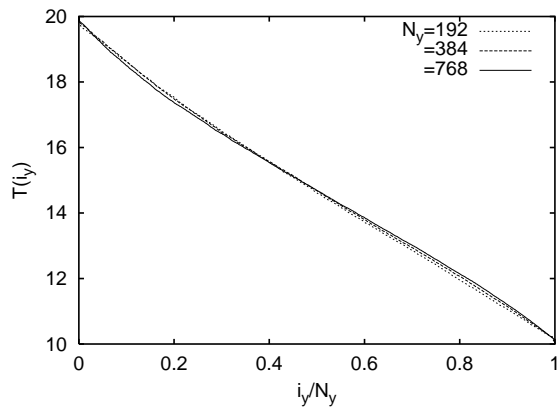}}
\caption{ Temperature profiles for 2D FPU-$\beta$ lattices with $N_x:N_y=1:2$. 
The sequences represent the results for the sizes $N_y= 192, 384$, and $768$. 
The temperatures of the heat baths at both ends were fixed to $(T_L, T_R)=(20.0, 10.0)$. 
The horizontal axis represents the position in the $y$-direction, scaled
by the system size $N_y$, and the vertical axis represents the local temperature. }
\label{fig:2dlimTemp} 
\end{figure}

\begin{figure}
\resizebox{0.90\linewidth}{!}{\includegraphics{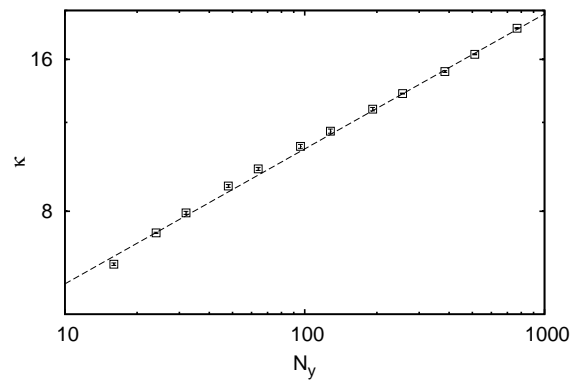}}
\caption{
System size dependence of the thermal conductivity for 2D
FPU-$\beta$ lattices plotted on a log-log scale. The dashed line
represents the result of a power-law fitting in the region $N_y\ge 128$,
yielding the result $\kappa (N_y)\sim N_y^{0.267(5)}$.}
\label{fig:2dlim} 
\end{figure}

We simulated a system consisting of a simple square
2D FPU-$\beta$ 
lattice whose Hamiltonian is of the same form as eq. (\ref{eq:3dhamil}). 
We fixed the aspect ratio to $N_x:N_y=1:2$, where the
$y$-axis is taken to be the direction of heat flow. For the size 
$384\times 768$, we waited for approximately $t_w\sim 5\times 10^5$
for the system to reach the nonequilibrium steady state. 

The temperature profile of the steady state is plotted in Fig. \ref{fig:2dlimTemp}.
It is seen that there are no temperature gaps near the walls.
Thus, we can define the thermal conductivity by eq. (\ref{eq:conductivity}). 

The system size dependence of the thermal conductivity is plotted in Fig. \ref{fig:2dlim}.
There, we see that the conductivity exhibits power-law divergence. 
Because the values of $N_y$ used here are much larger than the values of $N_z$ used
in the three-dimensional systems that we studied,
we can clearly distinguish the behavior here from logarithmic 
divergence. The data presented here are sufficient to conclude that 
the divergence is stronger than conventional logarithmic divergence.
Our estimation of the power-law exponent is
$\alpha\sim 0.268(3)$, obtained by fitting in the region $N_y\ge 128$.

\section{Summary and Conclusion}

Heat conduction in FPU-$\beta$ lattices was studied using nonequilibrium 
molecular dynamics simulations. The divergence of the thermal conductivity was
observed in simple cubic lattices up to the size $128\times 128\times 256$.
Such divergence was also observed in fcc lattice systems up to 
$192\times 192\times 193$, 
in fcc lattice systems with a natural length included in the nonlinear interaction, 
and in simple cubic lattice systems with randomly distributed particle masses with a 
variance of 20\% ($\lambda = 0.4$). 
Similar divergence for square lattice systems was also found for sizes up to $384\times 768$. 
These divergences are characterized by power-law dependences on the system size, 
and the power exponents do not appear to be unique, ranging from 0 (logarithmic) to $0.22$ 
for simple cubic lattice systems, 
and being approximately $0.5$ for fcc lattice systems. 
Convergence was observed in simple cubic lattice systems with randomly fixed particles.  

The present results imply that the long-time-tail decay exponent $-d/2$ of 
the energy flux autocorrelation function is not observed in the present 
systems with diverging thermal conductivity, although it has been observed
in particle systems with an order-of-magnitude smaller number of 
degrees of freedom\cite{Shimada,Murakami,Ogushi}. 
Of course, the size limitation is always an issue to consider in finite-system analysis. 
Decay with an exponent of $-3/2$ may appear in larger 3D
nonlinear lattices, and thermal conductivity may converge. 
For this reason, it would be worthwhile carrying out simulations on
larger lattices using more powerful computers. 
However, it should be remarked here that the $128\times 128\times 256$ lattice 
is already mesoscopic to macroscopic 
in the present technological sense: 
this lattice corresponds to insulators of size $50\times 50\times 100$nm$^3$ 
when the lattice constant is assumed to be of the scale of diamonds, for example. 
Experimental observations of heat transport in crystals of this scale should
be carried out. 

There remain several open problems in addition to the system size problem.
One is to elucidate the direction dependence of the heat transport;
in the present study, the heat flow was always in the [001] direction. 
Another is to make the interaction potential function more realistic.

\acknowledgements
We are grateful to Hisao Hayakawa, Masaharu Isobe,
Tomio Y. Petrosky, Akira Shimizu, and Akira Ueda for fruitful discussions. 
We also acknowledge Keiji Saito, Takashi Shimada, and Satoshi Yukawa 
for giving us ideas and assistance.
The numerical calculations were carried out on NEC SX8 at YITP in Kyoto University 
and Hitachi SR11000 at ISSP in the University of Tokyo. 
This work is partially supported by a Japan Society for the Promotion of 
Science Grant (No. 19340110).



\begin{thebibliography}{99}
\bibitem{Shimada} T. Shimada, T. Murakami, S. Yukawa, K. Saito, and N. Ito: J. Phys. Soc. Jpn. {\bf 69} (2000) 3150.
\bibitem{Murakami} T. Murakami, T. Shimada, S. Yukawa, and N. Ito: J. Phys. Soc. Jpn. {\bf 72} (2003) 1049.  
\bibitem{Ogushi} F. Ogushi, S. Yukawa, and N. Ito: J. Phys. Soc. Jpn. {\bf 74} (2005) 827.
\bibitem{Kaburaki} H. Kaburaki, J. Li, S. Yip, and H. Kimizuka, J. Appl. Phys. {\bf 102} (2007) 043514. 
\bibitem{Ishiwata}T. Ishiwata, T. Murakami, S. Yukawa, and N. Ito: Intern. J. Mod. Phys. C {\bf 15} (2004) 1413.
\bibitem{Yuge}T. Yuge, A. Shimizu, and N. Ito: J. Phys. Soc. Jpn. {\bf 74} (2005) 1895.
\bibitem{LepriR} S. Lepri, R. Livi, and A. Politi: Phys. Rep. {\bf 377} (2003) 1.
\bibitem{FPUbeta}  S. Lepri, R. Livi, and A. Politi: Phys. Rev. Lett. {\bf 78} (1997) 1896.
\bibitem{FPUalpha} S. Lepri: Eur. Phys. J. B {\bf 18} (2000) 441.
\bibitem{Hatano} T. Hatano: Phys. Rev. E {\bf 59} (1999) R1.
\bibitem{FPUother} H. Kaburaki and M. Machida: Phys. Lett. A {\bf 181} (1993) 85. 
\bibitem{AokiKusnezov} K. Aoki and D. Kusnezov: Phys. Rev. Lett. {\bf 86} (2001) 4029.
\bibitem{Shiba} H. Shiba, S. Yukawa, and N. Ito: J. Phys. Soc. Jpn. {\bf 75} (2006) 103001.
\bibitem{Lebowitz}  Z. Rieder, J. L. Lebowitz, and E. Lieb: J. Math. Phys. {\bf 8} (1967) 1073.
\bibitem{Peierls} R. E. Peierls: {\it Quantum Theory of Solids} (Oxford University Press, London, 1955).
\bibitem{hsg} A. Dhar, Phys. Rev. Lett. {\bf 86} (2001) 3554.
\bibitem{CasatiProsen} G. Casati and T. Prosen: Phys. Rev. E {\bf 67} (2003) 015203(R).
\bibitem{GK} M. S. Green: J. Chem. Phys. {\bf 22} 398 (1954).
\bibitem{GK2} R. Kubo: J. Phys. Soc. Jpn. {\bf 12} (1957) 570. 
\bibitem{GK3} R. Kubo, M. Yokota, and S. Nakajima: J. Phys. Soc. Jpn. {\bf 12} (1957) 1203.
\bibitem{LepriM} S. Lepri, R. Livi, and A. Politi: Europhys. Lett. {\bf 43} (1998) 271.
\bibitem{Narayan} O. Narayan and S. Ramaswamy: Phys. Rev. Lett. {\bf 89} (2002) 200601.
\bibitem{MaiNarayan} T. Mai and O. Narayan: Phys. Rev. E {\bf 73} (2006) 061202.
\bibitem{kinetic} A. Pereverzev: Phys. Rev. E {\bf 68} (2003) 056124.
\bibitem{longtail} Y. Pomeau and P. R\'esibois: Phys. Rep. {\bf 19} (1975) 63
\bibitem{Resibois} P. R\'esibois and and M. de Leener: {\it Classical Kinetic Theory of Fluids} (John Wiley \& Sons, New York, 1977).
\bibitem{Ernst} M. H. Ernst, E. H. Hauge, and J. M. J. Van Leeuwen: Phys. Lett. A{\bf 34} (1971) 419.
\bibitem{Ernst2} M. H. Ernst, E. H. Hauge, and J. M. J. Van Leeuwen: Phys. Rev. A {\bf 4} (1971) 2055. 
\bibitem{Ernst3} M. H. Ernst, E. H. Hauge, and J. M. J. Van Leeuwen: J. Stat. Phys. {\bf 15} (1976) 7.
\bibitem{Ernst4} M. H. Ernst, E. H. Hauge, and J. M. J. Van Leeuwen: J. Stat. Phys. {\bf 15} (1976) 23. 
\bibitem{Alder} B. J. Alder and T. E. Wainwright: Phys. Rev. Lett. {\bf 18} (1967) 988.
\bibitem{Alder2} B. J. Alder and T. E. Wainwright: Phys. Rev. A {\bf 1} (1970) 18.
\bibitem{longtail2} T. Yuge and A. Shimizu: J. Phys. Soc. Jpn. {\bf 76} (2007) 093001.
\bibitem{Nishino} T. H. Nishino: Prog. Theor. Phys. {\bf 118} (2007) 657. 
\bibitem{Fujii}  M. Fujii, X. Zhang, H. Xie, H. Ago, K. Takahashi, T. Ikuta, H. Abe, and T. Shimizu: Phys. Rev. Lett {\bf 95} (2005) 065502.
\bibitem{Shioya} H. Shioya, T. Iwai, D. Kondo, M. Nihei, and Y. Awano: Jpn. J. Appl. Phys. {\bf 46} (2007) 3139.
\bibitem{Maruyama} S. Maruyama, Physica B {\bf 323} (2002) 193.
\bibitem{Nose}S. Nos\'e: Prog. Theor. Phys. Suppl. {\bf 103} (1991) 1.
\bibitem{HDO1} D. N. Payton and W. M. Visscher: Phys. Rev. {\bf 156} (1967) 1032. 
\bibitem{HDO2} A. Casher and J. L. Lebowitz: J. Math. Phys. {\bf 12} (1971) 1701.
\bibitem{HDO3} A. J. O'Connor and J.  L. Lebowitz: J. Math. Phys. {\bf 15} (1974) 692.
\bibitem{HDO4} H. Matsuda and K. Ishii: Prog. Theor. Phys. Suppl. {\bf 45} (1970) 56.
\bibitem{LiD} B. Li, H. Zhao, and B. Hu: Phys. Rev. Lett. {\bf 86} (2001) 63.
\bibitem{mbd} L. Yang: Phys. Rev. Lett. {\bf 88} (2002) 094301.
\bibitem{LippiLivi} A. Lippi and R. Livi: J. Stat. Phys. {\bf 100} (2000) 1147.
\bibitem{Yang} L. Yang, P. Grassberger, and B. Hu: Phys. Rev. E {\bf 74} (2006) 062101.
\end{thebibliography}
\end{document}